\documentclass[a4paper,twoside,hyphens]{article}

\usepackage{epsfig}
\usepackage{subcaption}
\usepackage{calc}
\usepackage{amssymb}
\usepackage{amstext}
\usepackage{amsmath}
\usepackage{amsthm}
\usepackage{multicol}
\usepackage{pslatex}
\usepackage{apalike}
\usepackage{algorithm2e}
\usepackage[bottom]{footmisc}
\usepackage{xfrac}    
\usepackage{orcidlink}

\usepackage{SCITEPRESS}     

\usepackage{tikz}
\usepackage{pgfplots}
\pgfplotsset{compat=1.18}

\usepackage{hyperref}
\usepackage{wasysym}
\usepackage{subcaption}
\usepackage{booktabs}
\usepackage{xcolor}
\definecolor{tl_red}{rgb}{1,0.196,0.196}
\definecolor{tl_yellow}{rgb}{1,0.835,0.186}
\definecolor{tl_green}{rgb}{0.176,0.888,0.215}

\begin{document}

\title{Evaluating the Influence of Multi-Factor Authentication and Recovery Settings on the Security and Accessibility of User Accounts\thanks{Author version of the paper originally presented and published 
at \href{https://icissp.scitevents.org/?y=2024}{ICISSP 2024}. The accepted version can be found at \url{http://dx.doi.org/10.5220/0012319000003648} and its use is subject to the terms and conditions of SciTePress.}}

  \author{\authorname{Andre Büttner\orcidAuthor{0000-0002-0138-366X} and Nils Gruschka\orcidAuthor{0000-0001-7360-8314}}
  \affiliation{University of Oslo, Gaustadalléen 23B, 0373 Oslo, Norway}
 \email{\{andrbut,nilsgrus\}@ifi.uio.no}
 }


\abstract{Nowadays, most online services offer different authentication methods that users can set up for multi-factor authentication but also as a recovery method. This configuration must be done thoroughly to prevent an adversary's access while ensuring the legitimate user does not lose access to their account. This is particularly important for fundamental everyday services, where either failure would have severe consequences. Nevertheless, little research has been done on the authentication of actual users regarding security and the risk of being locked out of their accounts.
To foster research in this direction, this paper presents a study on the account settings of Google and Apple users. Considering the multi-factor authentication configuration and recovery options, we analyzed the account security and lock-out risks. Our results provide insights into the usage of multi-factor authentication in practice, show significant security differences between Google and Apple accounts, and reveal that many users would miss access to their accounts when losing a single authentication device.}

\onecolumn \maketitle \normalsize \setcounter{footnote}{0} \vfill

\section{Introduction}
Online services play an ever-increasingly important role in our digital world. We use them in many aspects of our private and business lives, like finance, transport, communication, entertainment, etc. Big tech companies---like Google and Apple---provide us with different services, including cloud storage, device management, payment, or single sign-on (SSO) to other online services. We highly depend on these services for more and more everyday tasks, and their availability and security are more crucial than ever.

Most online services require an account, and the user must authenticate to give proof of the ownership of their account. Authentication is still done mainly through passwords, although it is commonly known that passwords are a relatively weak method to protect a user account \cite{taneski2019systematic}. This has been confirmed in many studies, highlighting the problems with simple passwords and the reluctance to use password managers~\cite{hayashi2011diary,shen2016user,davis2022password}.
Therefore, multi-factor authentication (MFA) is usually offered to increase security and make compromising an account by a malicious user significantly harder. With MFA, the user needs to provide, in addition to the password, one\footnote{In this case, also called two-factor authentication (2FA). In this paper, we will generally use the term MFA.} or more authentication factors, such as a one-time password (OTP) from an authenticator app or sent via SMS or a security key. Despite the apparent advantage, many users consider MFA inconvenient and, therefore, refrain from activating it \cite{das2019mfa}. However, an increasing number of services make MFA mandatory and enforce configuring a second authentication factor. In addition, there is a movement towards passwordless authentication with the use of FIDO2 passkeys \cite{Google2022Passkeys}, but, at the time of writing, it is not widely adopted yet.

Suppose the MFA methods are not accessible, or users forget their password (or lose access to their password storage). In that case, most online services provide additional authentication factors, so-called account recovery mechanisms. Typical examples of these mechanisms are a link or code sent via email or SMS. However, such account recovery can be exploited if they are less secure than the main authentication, e.g., if a recovery email address is not protected sufficiently \cite{li2018email}. Also, a recovery factor might be bound to the same device as the main authentication. For example, the phone receiving the recovery SMS might be the same as the phone used for generating the MFA OTPs, which is certainly not an uncommon combination on a smartphone. Statistics show that smartphones often get stolen, damaged, or lost \cite{bitkom,mobile_theft_report}, which can (in addition to the financial loss) make connected accounts inaccessible. Many providers warn users if the authentication is insecurely configured (e.g., no MFA activated or weak password). Still, they can not derive which device the configured factors are bound to. Furthermore, little research was done on this issue.

To tackle this research question, we conducted an online survey with 185 test participants to study the authentication configuration of their Google and Apple accounts\footnote{We selected Google and Apple, as they are two of the largest online services and offer a large variety of authentication configuration options.}. 
To analyze the participants' responses, we extended and applied \textit{Account Access Graphs} (AAG) \cite{hammann2019user,pohn2022multi}, a new approach to evaluate MFA and recovery configurations. The analysis revealed insights into the account security and accessibility ``in the wild''. The main contributions of this paper are thereby as follows:
\begin{enumerate}
    \item An improved accessibility scoring for AAG models with higher practical significance.
     \item A security and accessibility analysis of the authentication setups by actual Apple and Google users.
\end{enumerate} 

The remainder of this paper is structured as follows. Section \ref{sec:related_work} presents related work on MFA and AAGs. In Section \ref{sec:aags}, the methodology of AAGs is described, as well as our security and accessibility scoring. Afterwards, we describe our study design in Section \ref{sec:study_design}. In Section \ref{sec:study_results}, we present our study results. Section \ref{sec:discussion} discusses the limitations, our new accessibility scoring, and the implications of our results on MFA. The paper is finally concluded in Section \ref{sec:conclusion}.

\section{Related Work}\label{sec:related_work}
A study from 2015 analyzed how widely MFA authentication is adopted among Gmail users \cite{petsas2015two}. Within their test set, 6.39\% of all users had an MFA method enabled, about 62\% had configured a phone for account recovery, and 17\% had provided a recovery email address. This gives a good picture of the adoption of MFA at the time of the study, but it might have changed significantly throughout the years. Their study is similar to ours regarding the collected data, i.e., enabled MFA and recovery methods. However, they obtained the data by using the \textit{Google Password Reminder} feature with leaked email lists while we conducted an anonymous survey with the participants' consent.

Much research was done on the authentication and account recovery procedures offered by online services. A study by Gavazzi et al. \cite{gavazzi2023study} investigated how popular websites support MFA and risk-based authentication (RBA). They found that among the 208 websites they tested, about 42\% provided MFA, and about 22\% applied RBA. Amft et al., for example, evaluated the account recovery of 1303 websites based on their documentation and tested the actual recovery procedure on 71 of these websites. By this, they discovered many insecure procedures, eventually allowing them to bypass MFA authentication through recovery and significant deviations between the documentation and the actual procedure \cite{amft2023we}. Another study analyzed usability flaws in account recovery when MFA is enabled on 78 popular websites \cite{gerlitz2023adventures}. Our paper, in contrast, focuses on which authentication and recovery methods provided by a service are adopted by users.

Other work has been done to compare different authentication methods based on security, privacy, and usability properties. The conclusion is that it is not feasible to find authentication schemes that perfectly satisfy all these properties \cite{ometov2018multi}. More recent research has conducted a formal analysis of multi-factor authentication methods provided by Google, revealing some weaknesses in their protocols \cite{jacomme2021extensive}. 

Furthermore, different studies exist that compare the usability of MFA methods.
For example, Reese et al. have looked at the perceived usability of five different MFA methods in terms of usability and their efficiency \cite{reese2019usability}. In another study, the time used on MFA methods has been compared in two universities, showing that this can take up a significant amount of working hours over a long time \cite{reynolds2020empirical}. 

Account Access Graphs \cite{hammann2019user,pohn2022multi} is a methodology to analyze authentication systems systematically. It allows the modeling of authentication and recovery methods of online services and the dependencies between different user accounts. Further, it is possible to define dedicated scoring schemes to evaluate various aspects of an account, such as security and accessibility. It was shown that this could even become a practical tool for users to increase awareness about their account use and to support them in making it more secure \cite{pohn2022multi}. Moreover, AAGs were used in a lab study to observe patterns in account setups \cite{hammann2022m}. The methodology used in this work mostly leans on the foundation provided in \cite{pohn2022multi}.

In this paper, we have enhanced the methodology of AAGs with an improved accessibility scoring scheme and applied it to actual user accounts. By automatically creating graphs of users, we show that analyzing user accounts with AAGs can be scalable for research on larger populations.

\section{Account Access Graphs}\label{sec:aags}
In an Account Access Graph as presented in \cite{pohn2022multi}, authentication methods and accounts are modeled as nodes in a graph. 
Account nodes are illustrated as rectangles, and authentication nodes as rounded rectangles. These nodes can be connected through intermediary nodes representing logical operators, including conjunctions '$\&$' and disjunctions '$|$', depicted as circles. For instance, MFA can be modeled as a conjunction of two or more authentication methods. Recovery or alternative authentication mechanisms are modeled as a disjunction. 
Graphs can be created for the entire authentication system of an online service or the account setup of a specific user and can be used for qualitative and quantitative analysis.

In this paper, we use AAGs to evaluate the security and accessibility of actual user accounts. The respective scoring schemes are described below. 

\subsection{Security Scoring}
\begin{table}[t!]
    \caption{Security scores assigned to different example authentication methods.}
    \label{tab:security_score}
    \centering
    \small
    \begin{tabular}{@{}llp{3.3cm}@{}}
    \toprule
           \textbf{Score} & \textbf{Category} &\textbf{Authentication Methods}  \\ 
    \midrule
          High &Hardware-based &  Security Key, Smart Card  \\
           Medium & Software-based & SMS Code, OTP Apps \\
           Low & Knowledge-based & Password, PIN \\
          \bottomrule
    \end{tabular}
\end{table}

As suggested in \cite{hammann2019user}, there are different possibilities to assess the security in an AAG. We adopted a scoring scheme inspired by the level of assurance as found in standards like NIST \cite{NIST_SP_800-63b} or the EU electronic identification and trust services (eIDAS) \cite{eIDAS_LoA}, which is a well-established method to evaluate authentication methods. Consequently, we chose to use an ordinal scale including \textit{low}, \textit{medium} and \textit{high}. 
An overview of how security values are assigned to different authentication methods is given in Table~\ref{tab:security_score}.

\begin{figure} [t]
    \centering
    \includegraphics[width=.9\linewidth]{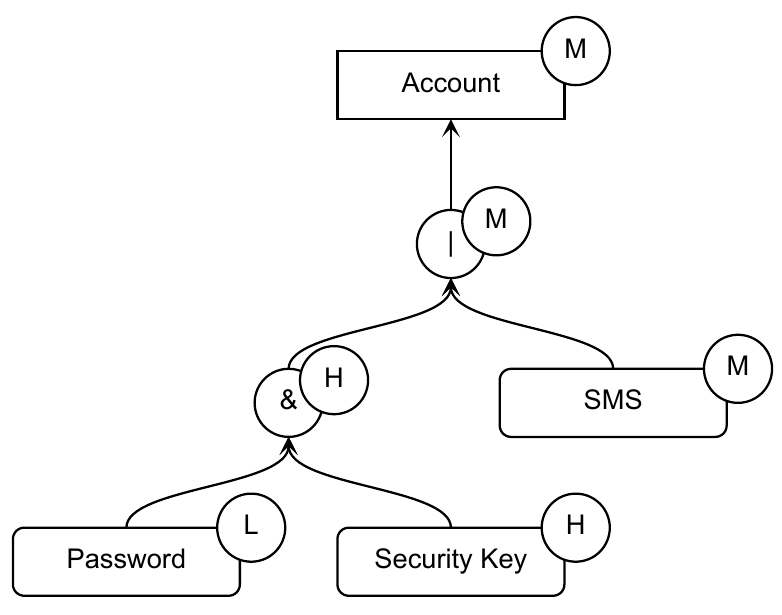}
    \caption{Example graph for showing how security scores are calculated. The scores are indicated as L (low), M (medium), and H (high).}
    \label{fig:security_example}
\end{figure}

The security scores of parent nodes are calculated as follows. For the logical conjunction, the score is calculated using the maximum score of the child nodes. This is because all child nodes must be accessed to access this node, so the highest score defines the security level. The score for a logical disjunction is calculated by the minimum score of the children, as accessing the weakest of the children is sufficient to access a node. Figure \ref{fig:security_example} shows an example for calculating security scores.

Some nodes in AAGs might represent other accounts, such as recovery email addresses. Ideally, the scores for these nodes should be derived from the respective AAGs specifically for that email account. However, since we could not consider all possible email accounts used, we opted for a worst-case calculation and assigned the value \textit{low} to those leaf nodes.

\subsection{Accessibility Scoring}\label{sec:accessibility_scoring}
Users might lose their credentials or a device required for MFA authentication and be locked out of their accounts. Account accessibility scores indicate how many options users have to access their accounts. Consequently, a low accessibility score implies a high risk and a high score means a low risk of losing access to an account.  

The accessibility scoring, as proposed in~\cite{pohn2022multi}, provides a first estimation of whether the accessibility of a user account is stronger or weaker compared to others. However, it is not precise enough to give practical conclusions. We, therefore, propose an improved accessibility scoring. 
\begin{figure} [t]
    \centering
    \includegraphics[width=.9\linewidth]{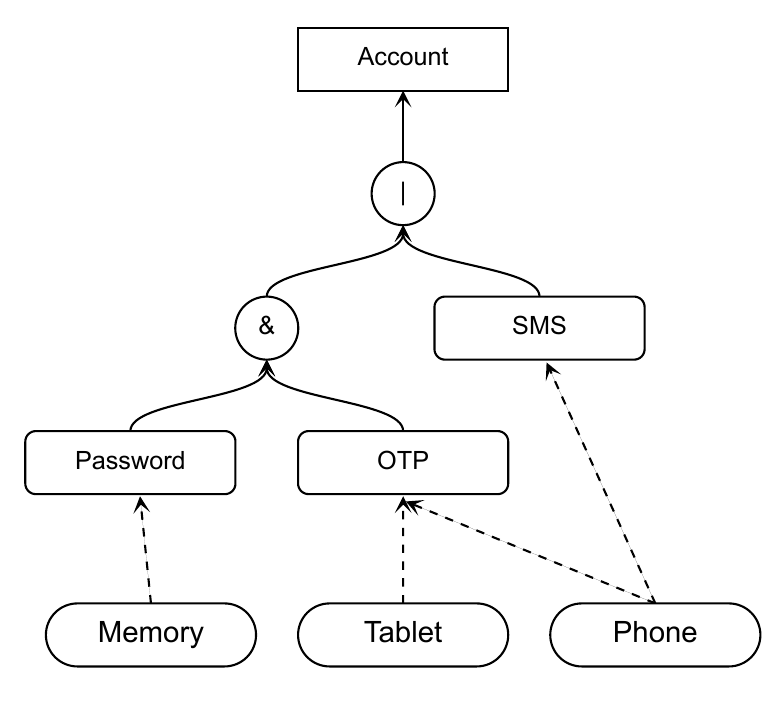}
    \caption{Example graph with access methods.}
    \label{fig:accessibility_example}
\end{figure}

The basic idea behind our improved scoring remains the same. All authentication methods have a physical ``representation'', e.g., passwords might be stored on a computer or remembered in memory, and SMS retrieval requires a phone to which it is sent. These elements are represented as \textit{access method} in the AAG graph and are connected to the respective authentication node.

In contrast to the other nodes in an AAG, access methods may have more than one parent node (as shown in Figure \ref{fig:accessibility_example}). If multiple different access methods can access an authentication method, it means that either one can access it. As a consequence, these access methods have a disjunctive relation. To assess the accessibility of a specific AAG, we use the logical formula based on the nodes in the graph.  

{\small
\begin{align} \label{eq:accessibility_example}
 &(\text{Memory} \wedge  (\text{Tablet} \vee \text{Phone})) \vee \text{Phone}\\
 &(\text{Memory} \wedge \text{Tablet}) \vee (\text{Memory} \wedge \text{Phone}) \vee \text{Phone}\\
&(\text{Memory} \wedge \text{Tablet}) \vee \text{Phone}
\end{align}
}

As an illustrating example, the equations above show the boolean terms describing the accessibility of the instance in Figure~\ref{fig:accessibility_example}. First, we derive the term from the leaf nodes of the graph and replace them with the respective access methods (1). We then transform the term (2) to get a complete list of possible combinations of access methods for accessing that account. Finally, we reduce the term so only those access methods remain that are at least required for accessing the account (3). This means that an account cannot be accessed if all of the device combinations in this term are inaccessible.

We derive a numerical score based on this reduced term for a simple comparison at a larger scale. This is done by counting the number of occurrences $n_{i}$ of each access method $i$ within the reduced term and assigning them the value $s_{i}=1/n_{i}$. The total accessibility score is calculated using the minimum of the conjunctive terms and the sum of the disjunctive terms. The resulting score $s_{acc}$ gives us the lower bound number of access methods one must lose to lose access completely. At the same time, $s_{acc}-1$ can be interpreted as the upper bound number of access methods that can be lost without any risk of being locked out of an account. For the given example, the accessibility score is calculated as follows:
\begin{align*}
     &s_{acc} = min(1, 1) + 1 = 2
\end{align*}
The accessibility score is $2$, and losing one access method is therefore not critical. However, losing two methods, e.g., the phone and tablet, can make the account inaccessible to the user.

\section{Study Design}
\label{sec:study_design}

This section presents our study design. In particular, the research questions to be answered by this study are formulated. Furthermore, we describe the study procedure, the AAG models created for Apple and Google, the user account survey, and some ethical considerations.

\subsection{Research Questions}
We designed a study to learn how users set up their Google and Apple accounts and how much they depend on their devices. One goal was to analyze how users access their passwords. Nowadays, password managers are becoming increasingly popular and are often even integrated into operating systems or browsers. This might affect the choice of passwords, i.e., whether they are easy to remember or rather created randomly and how accessible the password is. Furthermore, we wanted to check how users adopt MFA and recovery methods compared to earlier research \cite{petsas2015two}. In addition, we were interested in the general security of the users' accounts by considering the primary authentication and recovery methods. Finally, we wanted to assess how many devices users depend on. From this, the following research questions were derived:

\begin{itemize}
    \item \textbf{RQ1} How do the users access their passwords?
    \item \textbf{RQ2} Which MFA and recovery methods did the users enable?
    \item \textbf{RQ3} How secure are the account setups?
    \item \textbf{RQ4} How many access methods do the user accounts depend on?
\end{itemize}
\begin{figure*}[ht!]    
    \centering\includegraphics[width=.7\linewidth]{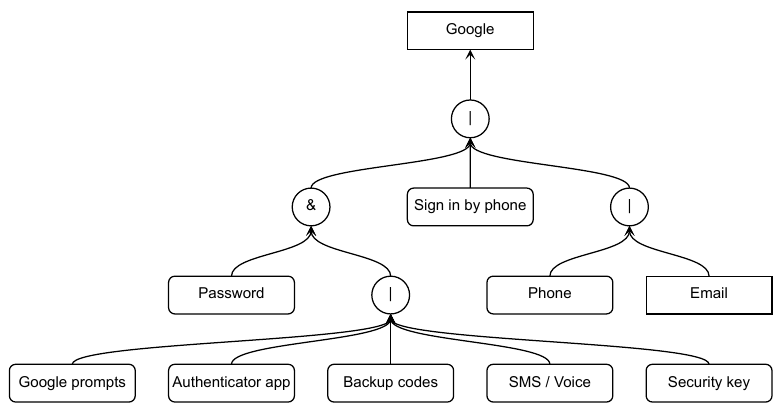}
    \caption{AAG for Google.}
    \label{fig:graph_google}
\end{figure*}
\subsection{Methodology}
For this study, we first created the models that describe the authentication systems for Google and Apple. We then created an online survey based on the authentication systems in which actual users were asked which authentication and recovery methods they use for their Google or Apple accounts and what means or devices they use to access them. After that, we generated user-specific AAG models based on the survey responses and calculated the respective security and accessibility scores. The general survey responses, as well as the security and accessibility scores, were finally analyzed concerning the research questions. 

A research tool and some scripts have been used to facilitate the creation and analysis of the AAGs\footnote{Our tools, survey questionnaires, anonymous participant data, and AAG files can be accessed at \url{https://github.com/Digital-Security-Lab/user-account-study-icissp2024} (Last accessed: 2023-12-06)}. 
The web application tool can create and visualize AAGs and calculate security and accessibility scores. Moreover, we created Python scripts to automatically convert the survey responses of the test participants into actual AAGs. These AAGs were then imported into the research tool and subsequently examined.

\subsection{General AAG Models}
The general AAG models for Google and Apple were created in January 2023. It was ensured that the AAGs were valid when the survey was conducted. This is important to consider as authentication systems change over time. For instance, since Spring 2023, Google has started to support FIDO2 passkeys as a passwordless alternative to its other authentication methods \cite{Google2023Passkeys}, which could not be covered in our study. The two AAG models used in this study are described in the following paragraphs.

\subsubsection*{Google}
To create an AAG model for Google accounts, the security settings have been examined. This is where users can change their passwords or enable other types of authentication. Note that figuring out all authentication methods requires experimenting because specific options are interdependent.
For Google accounts, the default primary authentication method is the password. In addition to that, it provides the option \textit{sign-in by phone} and several MFA methods. Google users can optionally set up a recovery email address or phone number for recovery.
The respective AAG model is shown in Figure~\ref{fig:graph_google}. In practice, Google is applying \textit{risk based authentication} (RBA) \cite{wiefling2019really}, which means that it might request different authentication factors depending on certain features of the user's client. This, however, can not be modeled by the AAGs because they currently only represent static models.

\subsubsection*{Apple}
\begin{figure}[t!]    
\centering
\includegraphics[width=\linewidth]{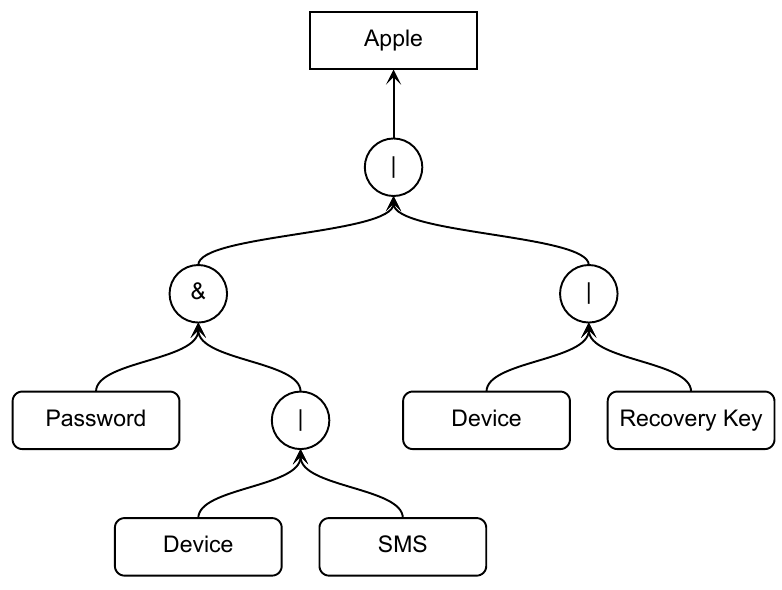}
    \caption{AAG for Apple.}
    \label{fig:graph_apple}
\end{figure}

The AAG model for Apple has been created based on the Apple ID security settings. As with Google, the primary authentication method here is a password.
However, users have fewer options to configure MFA in their Apple accounts. Secondary authentication factors can be either trusted phone numbers that are manually added by the user or trusted devices that are automatically configured when an Apple device is signed in to that Apple ID account. There is also the possibility of recovering accounts through customer service. This can not easily be modeled or evaluated using AAGs. Also, there is an option to set a recovery contact. These two options were excluded from the model. However, an Apple account can also be recovered using a trusted device. This option is implicitly enabled. Therefore, this has been added to the model.
Furthermore, a user can explicitly configure a printable recovery key. If this is selected, this is the only possible way to recover an Apple account. The final AAG model for the Apple account is shown in Figure~\ref{fig:graph_apple}.

\subsection{User Account Survey}
The surveys were created as two separate online forms, each adapted to Google or Apple account settings respectively. Both surveys consist of two main parts. 

First, the test participants had to make an enumerated list of the devices that they actively use. They should assign them to the categories of (1) \textit{phone}, (2) \textit{computer/laptop}, (3) \textit{tablet}, (4) \textit{smart watch} and (5) \textit{security key}. These devices were referred to in subsequent questions.

For the second part, the participants were supposed to log in to their Google or Apple accounts. 
Afterward, they had to specify by which means they could access their password, i.e., whether they could remember it, whether it was stored in a password manager, in a browser, on a device, or whether they wrote it down on paper. They could choose multiple options if applicable. After that, they were guided to specific account settings and asked about them.

In the case of Google, they were asked whether MFA methods were enabled and, if so, which ones. In addition, they had to choose the devices by which these methods could be accessed. For instance, if Google prompts were selected, they had to indicate which phones or tablets were configured for this method. If MFA was disabled, they were asked if \textit{sign-in by phone} was enabled and for which of the phones it was enabled. Last, they had to specify which recovery options were enabled, including email and phone. If a phone was selected, they had to choose the phone from their list of devices with the corresponding phone number. 

For Apple, participants had to indicate which devices were registered as trusted devices and which used a trusted phone number. Finally, the participants should state which recovery options were enabled in their Apple accounts.

\subsection{Ethical Considerations}
In the survey, the test participants were asked about what means they use to authenticate themselves to their Google or Apple accounts. This could generally be misused to exploit weak account configurations. However, the survey was completely anonymous, and no personal data was collected that could identify any of the participants by any means. Also, no other sensitive information, such as passwords or other secrets, was collected. The study was conducted in compliance with our university's research ethics guidelines. The test participants were thoroughly informed about the procedure and fairly compensated for their time. They could stop participating in the study at any time.

\section{Study Results}\label{sec:study_results}
After several preliminary test runs, the survey was conducted in mid-January 2023, with test participants acquired using Prolific \cite{Prolific2023}. There were 94 submissions for Google (age 18-70, MV 34.89, SD 10.25) and 91 for Apple (age 19-67, MV 32.77, SD 9.7). In both cases, about half of the participants were residents of the USA and the other half of Germany. Below, we describe the study results concerning the research questions.

\subsection{Password Access (RQ1)}
To answer RQ1, we assessed how people access their passwords. A summary of the responses from our survey is given in Table~\ref{tab:password_usage}. Most participants still tend to use a password they can remember. This applies to both Google and Apple. Yet, many also use a password manager or the password storage functionality of a browser or device. According to our results, less than 10\% of the participants have written it down on paper. It can be further observed that about one-third of the Apple test participants use more than one method to store their password, which applies to almost 48\% of the Google participants.
\begin{table}[t!]
    \caption{Responses of test participants about their password usage.}
    \label{tab:password_usage}
    \centering
    \small
    \begin{tabular}{p{3.8cm}p{1.2cm}p{1.2cm}}
    \toprule
          \textbf{Password access} & \textbf{Apple} \newline ($n=91$)  &   
          \textbf{Google} \newline($n=94$) \\\midrule
         I can remember it & 64  & 64  \\
          Password manager &  28 & 37  \\
          Stored by browser/device & 28 & 46  \\ 
          I wrote it down on paper & 9 & 7   \\
          \bottomrule
    \end{tabular}
\end{table}

\subsection{Adoption of MFA and Recovery Methods (RQ2)}
In RQ2, we wanted to analyze to what extent the users have adopted MFA and recovery methods. Apple and Google are very different in their approach to MFA. Apple devices are implicitly configured as MFA methods for accounts they are signed in. Therefore, Apple users most likely have MFA enabled automatically, so we did not examine this for Apple in more detail. 

Google users have several different authentication and recovery methods that need to be configured manually. The usage of each method is shown in Table~\ref{tab:google_auth_methods}. According to our results, 68\% of the Google test participants have at least one MFA method enabled. In the previously mentioned study from 2015, only less than 7\% had MFA enabled \cite{petsas2015two}. Furthermore, in 2018, a Google researcher stated that by that time, only 10\% of Gmail users had set up MFA \cite{TheAnatomy2018Usenix}. Compared to that, there is a significant increase in our test results. One must keep in mind that our sample size of 94 Google test participants is relatively small. Yet, an increase was expected since Google started to enroll MFA automatically for its users as announced in 2021 \cite{risher2021}. The most common secondary authentication method is voice or text message. Many of the participants also appear to use Google prompts and authenticator apps. Backup codes and security keys, in contrast, are used only rarely.

Regarding recovery methods, there also seems to be a wider adoption of a recovery email address (80\%) and a slightly increased use of the phone as a recovery method (68\%) compared to the results in \cite{petsas2015two}.

\begin{table}[t!]
    \caption{Frequency of authentication and recovery methods used in the participants' Google accounts ($n=94$).}
    \label{tab:google_auth_methods}
    \centering
    \small
    \begin{tabular}{p{3.5cm} p{1cm}p{1.7cm}}
    \toprule
    \textbf{Authentication method} & \textbf{MFA} & \textbf{Frequency}  \\
    \midrule
          Password-only & \Circle & 22  \\
          Sign-in by phone & \Circle & 8 \\
          Google prompts & \CIRCLE & 26 \\ 
          Authenticator app & \CIRCLE & 15 \\ 
          Backup codes & \CIRCLE & 8 \\ 
          Voice or text message & \CIRCLE & 38 \\
          Security key & \CIRCLE & 2\\
          Recovery phone & n/a & 64\\
          Recovery email & n/a & 76\\
          \bottomrule
          
    \end{tabular}
\end{table}

\subsection{Account Security (RQ3)}
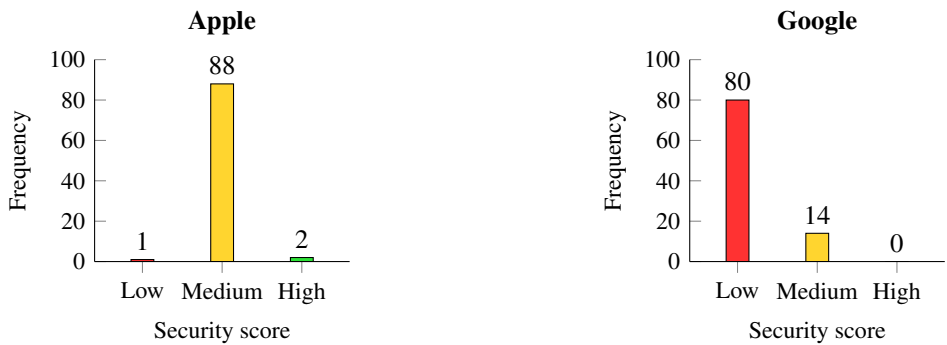
\begin{figure*}[ht!]
\begin{subfigure}{.49\linewidth}
\begin{tikzpicture}
  \centering
  \begin{axis}[
         ybar, axis on top,
        bar width=0.3cm,
        height=.55\linewidth,
        major grid style={draw=white},
        ymin=0, ymax=100,
        axis x line*=bottom,
        axis y line*=left,
        tick label style={font=\small},
        label style={font=\small},
        enlarge x limits=0.3,
        title={\textbf{Apple}},
        legend style={
            at={(0.5,-0.35)},
            anchor=north,
            legend columns=-1,
            /tikz/every even column/.append style={column sep=0.2cm},
            nodes={anchor=west,scale=0.8,},
        },
        ylabel={Frequency},
        xlabel={Security score},
       xtick=data,
       every axis plot/.append style={
          bar shift=0pt,
        },
        symbolic x coords={Low,Medium,High},
    ]
    \addplot [draw=black, fill=tl_red!100] coordinates { (Low,0) (Medium,0) (High,0)};
    \addplot [draw=black, fill=tl_red!100, nodes near coords={\pgfmathprintnumber[precision=0]{\pgfplotspointmeta}}] coordinates { (Low,1)};
    \addplot [draw=black, fill=tl_yellow!100, nodes near coords={\pgfmathprintnumber[precision=0]{\pgfplotspointmeta}}] coordinates { (Medium, 88) };
    \addplot [draw=black, fill=tl_green!100, nodes near coords={\pgfmathprintnumber[precision=0]{\pgfplotspointmeta}}] coordinates {  (High,2) };
  \end{axis}
  \end{tikzpicture}
\end{subfigure}
\begin{subfigure}{.49\linewidth}
\begin{tikzpicture}
  \centering
  \begin{axis}[
         ybar, axis on top,
        bar width=0.3cm,
        height=.55\linewidth,
        major grid style={draw=white},
        ymin=0, ymax=100,
        axis x line*=bottom,
        axis y line*=left,
        enlarge x limits=0.3,
        title={\textbf{Google}},
        tick label style={font=\small},
        label style={font=\small},
        legend style={
            at={(0.5,-0.35)},
            anchor=north,
            legend columns=-1,
            /tikz/every even column/.append style={column sep=0.2cm},
            nodes={anchor=west,scale=0.8,},
        },
        ylabel={Frequency},
        xlabel={Security score},
        symbolic x coords={
           Low, Medium, High},
       every axis plot/.append style={
          bar shift=0pt,
        },
       xtick=data
    ]
      \addplot [draw=black, fill=tl_red!100] coordinates { (Low,0) (Medium,0) (High,0)};
    \addplot [draw=black, fill=tl_red!100, nodes near coords={\pgfmathprintnumber[precision=0]{\pgfplotspointmeta}}] coordinates { (Low,80)};
    \addplot [draw=black, fill=tl_yellow!100, nodes near coords={\pgfmathprintnumber[precision=0]{\pgfplotspointmeta}}] coordinates { (Medium, 14) };
    \addplot [draw=black, fill=tl_green!100, nodes near coords={\pgfmathprintnumber[precision=0]{\pgfplotspointmeta}}] coordinates {  (High,0) };
  \end{axis}
  \end{tikzpicture}
\end{subfigure}
    \caption{Histogram over security scores of the participants' Apple and Google accounts.}
    \label{fig:histograms_security}
\end{figure*}

RQ3 was addressed by assessing the security scores of the AAGs created for all test participants. In general, it was observed that the recovery authentication methods of Google often decrease the security of an account, even if a user intended to use more secure authentication methods like the security key. This was particularly the case because most test participants used a recovery email. Nevertheless, it must be kept in mind that the security score for email was set to \textit{low} because the accurate score could not be assessed within this study. In reality, the score might be higher depending on the email provider and account setup. However, as mentioned before, it was shown that recovery emails are often a weakness in an account setup \cite{li2018email}.
The frequency of each security score for Google and Apple is summarized in~Figure~\ref{fig:histograms_security}.  

One can observe that there is no Google account with a total security score \textit{high}. The only way to achieve this would be to disable all recovery methods and only enable a security key as a second authentication factor. Two test participants have not enabled recovery methods but used Google prompts as the second authentication factor, with a score of \textit{medium} within our scoring scheme. Most Google accounts have a total score of \textit{low}. This is mainly because most test participants have enabled email as a recovery method. The lesson learned is that Google account security in practice often depends on the security of recovery email accounts. 

Two Apple users are rated with the score \textit{high}. This is because they have only set up a \textit{trusted device} but no \textit{trusted phone number}. Most Apple accounts have a score of \textit{medium}. The main reason here is likely that most Apple users use their account in connection with at least one Apple device, which, as mentioned before, automatically becomes a second authentication factor.

\subsection{Account Accessibility (RQ4)}
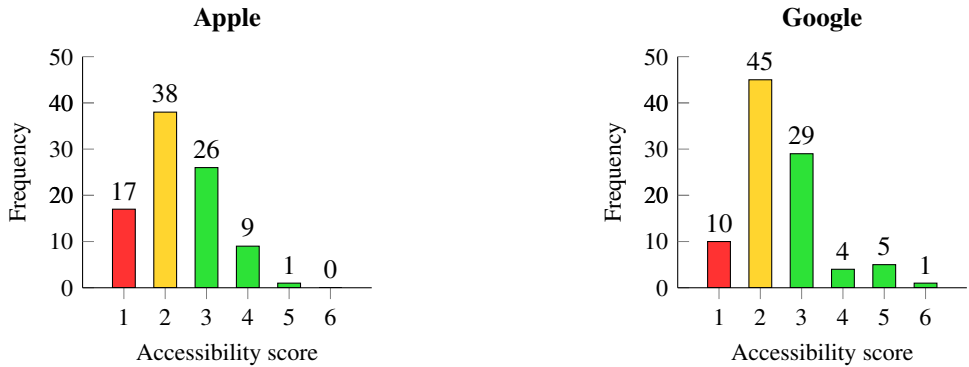
\begin{figure*}[t!]
\begin{subfigure}{0.49\linewidth}
\begin{tikzpicture}
\centering
  \begin{axis}[
         ybar, axis on top,
        bar width=0.3cm,
        height=.6\linewidth,
        major grid style={draw=white},
        ymin=0, ymax=50,
        axis x line*=bottom,
        axis y line*=left,
        enlarge x limits=0.2,
        tick label style={font=\small},
        label style={font=\small},
        title={\textbf{Apple}},
     extra y ticks = {10,20,30,40,50},
        legend style={
            at={(0.5,-0.35)},
            anchor=north,
            legend columns=-1,
            /tikz/every even column/.append style={column sep=0.2cm},
            nodes={anchor=west,scale=0.8,},
        },
        ylabel={Frequency},
        xlabel={Accessibility score},
        symbolic x coords={
           1,2,3,4,5,6},
            every axis plot/.append style={
          bar shift=0pt,
        },
       xtick=data
    ]
      \addplot [draw=black, fill=tl_red!100,] coordinates { (1,0) (2,0) (3,0) (4,0) (5,0) (6,0)};
    \addplot [draw=black, fill=tl_red!100, nodes near coords={\pgfmathprintnumber[precision=0]{\pgfplotspointmeta}}] coordinates { (1, 17)  };
    \addplot [draw=black, fill=tl_yellow!100, nodes near coords={\pgfmathprintnumber[precision=0]{\pgfplotspointmeta}}] coordinates { (2,38) };
    \addplot [draw=black, fill=tl_green!100, nodes near coords={\pgfmathprintnumber[precision=0]{\pgfplotspointmeta}}] coordinates {  (3, 26) (4,9) (5, 1) (6,0)};
  \end{axis}
\end{tikzpicture}
\end{subfigure}
\begin{subfigure}{0.49\linewidth}
\begin{tikzpicture}
\centering
  \begin{axis}[
         ybar, axis on top,
        bar width=0.3cm,
        height=.6\linewidth,
        major grid style={draw=white},
        ymin=0, ymax=50,
        axis x line*=bottom,
        axis y line*=left,
        enlarge x limits=0.2,
        extra y ticks = {10,20,30,40,50},
        tick label style={font=\small},
        label style={font=\small},
        title={\textbf{Google}},
        legend style={
            at={(0.5,-0.35)},
            anchor=north,
            legend columns=-1,
            /tikz/every even column/.append style={column sep=0.2cm},
            nodes={anchor=west,scale=0.8,},
        },
        ylabel={Frequency},
        xlabel={Accessibility score},
        symbolic x coords={
           1,2,3,4,5,6},
            every axis plot/.append style={
          bar shift=0pt,
        },
       xtick=data
    ]
    
     \addplot [draw=black, fill=tl_red!100] coordinates { (1,0) (2,0) (3,0) (4,0) (5,0) (6,0)};
    \addplot [draw=black, fill=tl_red!100, nodes near coords={\pgfmathprintnumber[precision=0]{\pgfplotspointmeta}}] coordinates { (1, 10)  };
    \addplot [draw=black, fill=tl_yellow!100, nodes near coords={\pgfmathprintnumber[precision=0]{\pgfplotspointmeta}}] coordinates { (2,45) };
    \addplot [draw=black, fill=tl_green!100, nodes near coords={\pgfmathprintnumber[precision=0]{\pgfplotspointmeta}}] coordinates {  (3, 29) (4,4) (5, 5) (6,1) };
    
  \end{axis}
\end{tikzpicture}
\end{subfigure}
    \caption{Histogram over accessibility scores of the participants' Apple and Google accounts.}
    \label{fig:histograms_accessibility}
\end{figure*}

We answered RQ4 by analyzing the accessibility scores for each test participant's Google and Apple accounts. The scores were derived from the respective AAGs, including the access methods used by the test participants. In Figure~\ref{fig:histograms_accessibility}, it can be observed that the range of accessibility scores is between 1 and 5 for Apple and between 1 and 6 for Google. 

As described in Section~\ref{sec:accessibility_scoring}, an accessibility score equal to 1 indicates that an account can become inaccessible if one of the access methods is unavailable, and we, therefore, analyzed these setups in more detail. Within the test sample, ten of the Google accounts have an accessibility score equal to 1. For Apple, this applies to seventeen accounts. After manual analysis, it was found that in sixteen cases for Apple and all ten cases for Google, the primary phone was the key access method that, if lost, could cause an account lockout. This is reasonable since smartphones play a crucial role in our lives today. The Google account of one test participant was even dependent on the availability of both a memorized password and a phone for Google prompts. 

Beyond that, most test participants had accessibility scores of 2 or higher. This means that the majority's risk of being locked out of their account is relatively low.

\section{Discussion}\label{sec:discussion}
In this section, we discuss the limitations of the conducted study, reflect on the new accessibility scoring, and, finally, discuss what conclusions can be drawn by our research for MFA in practice and how our study might contribute to future improvements.

\subsection{Limitations}
First of all, the study was conducted with a relatively low sample size of 91 Apple users and 94 Google users. Hence, the study is rather an initial study, and the results must be interpreted with that in mind. Yet, the results give us an idea of how many possible account setups are used in practice and what potential accessibility risks exist. Also, the results only reflect a snapshot of when the study was conducted, as authentication systems change with time.

Another limitation is the simplification of the Google model, where email accounts have been assigned a low security score. In practice, this will differ for each account. However, it requires a rather sophisticated survey covering all possible email account setups. 

Moreover, as mentioned earlier, RBA has not yet been considered in AAGs because the models are currently static. This might influence both security and accessibility. For that, a better understanding of RBA is required, and future work should look into how dynamic AAGs may be modeled and evaluated.

\subsection{Accessibility Scoring}\label{sec:discussion:accessibility_scoring}
An essential contribution of this work is the new accessibility scoring described in Section \ref{sec:accessibility_scoring}. In contrast to the previously proposed method \cite{pohn2022multi}, where no transformation or reduction was conducted after assessing the first Boolean term, our approach has a more practical meaning since it shows how many access methods a user depends on. For the example shown in Figure~\ref{fig:accessibility_example}, the score as in \cite{pohn2022multi} would result in an accessibility score of $1.5$, while our method resulted in a score of $2$. 

A numerical score does not provide context about the devices. Therefore, it could be beneficial to parse the logical formula into a human-readable description with more context about the actual devices. For instance, the final term from the example in Figure~\ref{fig:accessibility_example} can be compiled into a description like ``Access to \textit{Account} might be lost when losing both \textit{Phone} and \textit{Tablet}, or losing your \textit{Phone} and forgetting your password''. In practice, this could help inform online users more effectively of the consequences of losing a particular access method and motivate them to set up an alternative access method for authentication or recovery.

\subsection{MFA in Practice}
Our study shows that several users depend solely on their primary phones. For online services, it is challenging to assess the risk of their users being locked out of their account because the services can not know if, e.g., an authenticator app is installed on the same device that is used for SMS as a second authentication factor, or that has stored a user's password. We found that, especially for Apple, seventeen test participants might lose access to their accounts if they lose their phones. For Google, this applies to ten users. Given that, this also means that having access to the users' phones is enough to access an account. From a user perspective, one could argue that the term \textit{multi-factor} authentication does not always apply, even though users have enabled multiple authentication methods in their accounts. The likelihood of a phone being compromised is relatively low, as it usually requires local authentication through a PIN or biometrics or rather sophisticated malware. It is still a scenario that must be considered, especially by users with high-security requirements, e.g., due to their profession.

Also, we found that MFA and recovery are often linked to the same device. Beyond this study, we observed that, e.g., Facebook prohibits using the same phone number for MFA and account recovery, which is a step in the right direction. Surprisingly, other applications, such as LinkedIn, automatically enable both options when setting up a phone number. Web applications should encourage users to use different devices for different authentication methods. Therefore, it might be helpful to have a tool for users to keep an overview of their account and device dependencies, as suggested in \cite{pohn2022multi}.

It is generally not very obvious how accessible an account is. This depends on how people set up their accounts and how devices and password access are linked. AAGs can help assess this and encourage service providers and users themselves to improve the security and accessibility of online accounts. Therefore, we suggest the integration of AAGs into consumer tools and online services. Likewise, AAGs can be a powerful tool in an enterprise context for an administrator to get an overview of the account security and accessibility of employees.

\section{Conclusion}\label{sec:conclusion}
In this paper, we have extended the methodology of Account Access Graphs with a new accessibility scoring scheme. Moreover, we have developed and conducted a study in which we analyzed several aspects of Apple and Google user accounts, such as their MFA adoption and their security and accessibility. Within our security scoring scheme, Apple accounts turned out to be more secure, likely due to the nature of Apple devices being implicitly configured for multi-factor authentication. For Google, we found that several users still have not enabled any second authentication factor. One of the most important findings concerning accessibility was that several Google test participants and even more Apple test participants entirely depended on their phones. Even though this was not the majority, several users risk losing access to their user accounts. All in all, our study has demonstrated that AAGs can contribute to a deeper understanding of authentication methods. 

We will pursue this research direction in future work. Similar studies may be conducted for other web services to see how secure and accessible their users' account setups are. Furthermore, it can be investigated how service providers can use the knowledge gained to improve their authentication systems concerning security and accessibility. Finally, dynamic AAG models may be developed for analyzing accounts of online services that apply risk-based authentication.

\bibliographystyle{apalike}
{\small
\bibliography{references}}

\end{document}